\documentclass[%
 reprint,
 amsmath,amssymb,
 aps,
prb,
]{revtex4-1}

\usepackage{graphicx}
\usepackage{dcolumn}
\usepackage{bm}

\begin{document}

\preprint{APS/123-QED}

\title{Electronic properties of one-dimensional nanostructures of the Bi$_2$Se$_3$ topological insulator}

\author{Naunidh Virk}
\author{Gabriel Aut\`{e}s}
\author{Oleg V. Yazyev}%
 \email{E-mail: oleg.yazyev@epfl.ch}
\affiliation{Institute of Physics, Ecole Polytechnique F\'{e}d\'{e}rale de Lausanne (EPFL), CH-1015 Lausanne, Switzerland 
}%

\date{\today}

\begin{abstract}
We theoretically study the electronic structure and spin properties of one-dimensional nanostructures of the prototypical bulk topological insulator Bi$_2$Se$_3$.  
Realistic models of experimentally observed Bi$_2$Se$_3$ nanowires and nanoribbons are considered using the tight-binding method.
At low energies, the band structures are composed of a series of evenly spaced degenerate sub-bands resulting from circumferential confinement of the topological surface states. The direct band gaps due to the non-trivial $\pi$ Berry phase show a clear dependence on the circumference. The spin-momentum locking of the topological surface states results in a pronounced 2$\pi$ spin rotation around the circumference with the degree of spin polarization dependent on the the momentum along the nanostructure. Overall, the band structures and spin textures are more complicated for nanoribbons, which expose two distinct facets.
The effects of reduced dimensionality are rationalized with the help of a simple model that considers circumferential quantization of the topological surface states. 
Furthermore, the surface spin density induced by electric current along the nanostructure shows a pronounced oscillatory dependence on the charge-carrier energy, which can be exploited in spintronics applications.
\end{abstract}

\maketitle


\section{\label{sec:level1}Introduction}

Three-dimensional (3D) Z$_{2}$ topological insulators (TIs), with layered bismuth chalcogenides being the most common representatives of this family of materials, are characterized by the presence of topologically protected surface states (SSs) with a non-degenerate Dirac cone band dispersion.\cite{Zhang2009,Xia2009,Hsieh2009,Chen2009} A hallmark feature of such topological SSs is a helical spin polarization, whereby the electron spin is locked perpendicular to momentum.\cite{Ando2013,Qi2011,Hasan2010} The characterization of the surface-state Dirac cone band dispersion, and the confirmation of its spin helicity, has been well established through surface sensitive techniques such as angle-resolved photoemission spectroscopy (ARPES)\cite{Xia2009,Hsieh2009,Chen2009} and scanning tunneling microscopy (STM).\cite{Zhang2009a,Alpichshev2010,Beidenkopf2011} However, directly accessing and manipulating the topological properties of SSs through transport measurements has been hampered by an overriding bulk contribution to the electronic transport.\cite{Kong2011,Navratil2004,Kim2012a,Analytis2010} Low-dimensional nanostructures offer an efficient solution to overcoming this problem, as the large surface-to-volume ratio enhances SS contribution.\cite{Wang2012a,Xiu2011,Steinberg2010,Kong2011d,Hong2012,Hao2012,Kong2013} Commonly synthesized morphologies of bismuth chalcogenide nanostructures include nanowires, nanoribbons and nanoplatelets.\cite{Kong2009,Kong2010,Zou2014,Nowka2015,Peng2010,Li2012} Intriguingly, the morphology of nanowires and nanoribbons allow the TI SS spin helicity to be directly probed, as evidenced by the recent measurements of Aharonov-Bohm (AB) oscillations\cite{Aharonov1959a,Washburn1986,Aronov1987} originating from topological SSs.\cite{Hong2014,Cho2015,Jauregui2016} Thus, nanostructures show great promise in harnessing the helical properties of TI SS for applications in future electronic technologies such as spintronics\cite{Pesin2012,Mellnik2014,Li2014} and topological quantum computing.\cite{Stern2013}

The above-mentioned applications require a detailed understanding of how nanostructure morphology and finite-size effects affect the electronic structure, in particular how they relate to the topological phase. As bismuth chalcogenides TIs are layered materials, comprised of quintuple layers (QLs) held together by weak van der Waals (vdW) interactions, they possess an inherent cleavage plane, which corresponds to the extensively investigated (0001) surface. However, the reduced dimensionality of the nanostructures implies that surfaces other than the (0001) are inevitably present. Hence, as a primary step it is necessary to understand these high-index surfaces. In a previous work,\cite{Virk2016} we employed first-principles calculations for determining the structure and chemical composition of stable high-index surfaces of bismuth chalcogenide TIs. Moreover, we found that the band dispersion and spin texture of topological SSs are dependent on surface orientation and local chemical composition, in corroboration with associated experimental\cite{Xu2013,He2013} and theoretical\cite{Moon2011} studies.

In this paper, we theoretically investigate realistic models of one-dimensional (1D) nanostructures of TIs using a tight-binding approximation formalism. We  focus on bismuth selenide Bi$_{2}$Se$_{3}$ as the most common TI. In accordance with experiments and our previous work, we construct (see Section~\ref{Meth}) two distinct types of morphologies: hexagonal cross-section nanowires (NWs) and nanoribbons (NRs). Tight-binding calculations are utilized to investigate how finite-size effects and reduced dimensionality affect the electronic structure features derived from the topological SSs of each nanostructure. The dependence of the surface-state band gap on nanostructure dimensions is determined from band structure calculations. The effects of dimensionality reduction on the spin helicity are explored as a result of transitioning from a Dirac cone, associated with a 2D surface, to quantized 1D sub-bands, associated with a nanostructure.

\section{Methodology}\label{Meth}

\begin{figure*}
  \centering
  \includegraphics[width=0.7\textwidth]{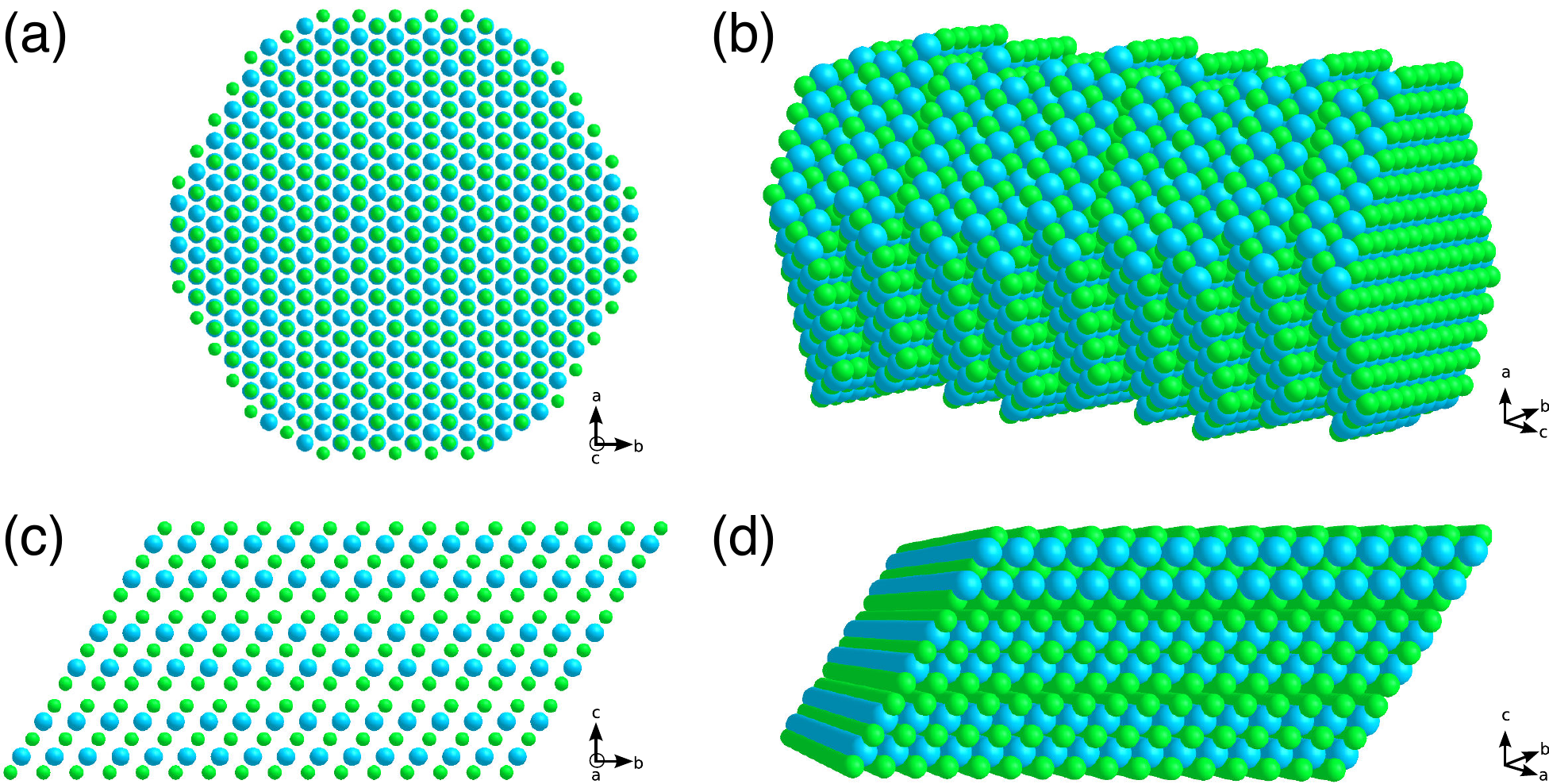}
  \caption{Examples of atomistic models of a Bi$_{2}$Se$_{3}$ hexagonal nanowire (perimeter 30~nm) and a nanoribbon (perimeter 19~nm). (a) Profile view showing cross-sectional area of the NW; $ab$ plane corresponds to the (0001) facet. (b) Perspective view showing QL stacking along the NW axis (i.e. along lattice vector $c$). (c) Profile view showing cross-sectional area of the NR, $cb$ plane corresponds to the (2$\bar{1}\bar{1}$0) facet, and NR axis is parallel to the direction of lattice vector $a$. (d) Perspective view showing axial direction of NR axis along $a$. In all plots blue spheres correspond to Bi atoms, whilst green spheres to Se atoms.}
  \label{fig:Bi2Se3_HexNW_NR_Models}
\end{figure*}

Tight-binding (TB) calculations were performed on two distinct morphologies of Bi$_{2}$Se$_{3}$ nanostructures: hexagonal nanowires and nanoribbons (Fig.~\ref{fig:Bi2Se3_HexNW_NR_Models}). 
We adopted TB parametrization of Bi$_{2}$Se$_{3}$ fitted to the results of first-principles calculations from Ref.~\onlinecite{Kobayashi2011b}. This model has already been utilized in the studies of the electronic structure of TI nanostructures, {\it e.g.} in Ref.~\onlinecite{Pertsova16}. In this TB model, an $sp^3$ atomic orbital basis set is used for both Bi and Se. Hopping integrals describing interactions between neighbor atoms in the same atomic plane and between atoms in neighboring and second-neighbor planes are taken into account. Spin-orbit coupling is included via the $p$-orbital spin-orbit matrix elements.

The choice of the TB method was imposed by the size of the NW and NR systems under investigation, where the largest NW and NR models contain $7005$  and $2160$ atoms, respectively. The size of the Hamiltonian matrices corresponding to these models largely exceeds $10^4$.  
The use of first-principles methods such as DFT would be prohibitively expensive. One potential issue with the TB model utilised is that surface and edge potential effects are not accounted for. However, all the NR systems being investigated have surfaces which either correspond to the (0001) surface, or are defined by a stoichiometric QL edge termination, where our previous work \cite{Virk2016} has shown almost no edge relaxation takes place. Consequently, we believe that the exclusion of these effects in the TB model does not impact upon any conclusions that can be drawn from our results. Eigenvalues and eigenvectors close to the Fermi level were obtained using a Jacobi-Davidson method implemented for large sparse symmetric matrices.\cite{jadamilu}

In order to describe the atomistic models used in our work we will refer to crystallographic planes and lattice directions in the four-index Miller-Bravais notation. The circumference of the nanostructures is expressed in terms of the perimeter of their cross-sections.
For hexagonal nanowires (Fig.~\ref{fig:Bi2Se3_HexNW_NR_Models}a,b), the axial direction of the NW is the [0001] direction ($c$ in Fig.\ref{fig:Bi2Se3_HexNW_NR_Models}a). This is parallel to the direction along which QLs stack (Fig.~\ref{fig:Bi2Se3_HexNW_NR_Models}b), and is also the experimental growth direction of synthesized Bi$_{2}$Se$_{3}$ NWs.\cite{Zou2014,Kong2009} Each QL in the NW is defined by six hexagonal edges (Fig.~\ref{fig:Bi2Se3_HexNW_NR_Models}a). Our recent work on Bi$_{2}$Se$_{3}$ high-index surfaces elucidated that the most stable QL edge termination, in a wide range of realistic chemical potential values, is a stoichiometric termination oriented along the [2$\bar{1}\bar{1}$0] direction.\cite{Virk2016} 
Consequently, the hexagonal NWs were constructed such that their facets correspond to the stoichiometric [2$\bar{1}\bar{1}$0] QL edge termination. A perspective view of two hexagonal NW edge surfaces is shown in Fig.~\ref{fig:Bi2Se3_HexNW_NR_Models}b. The QL stacking along the [0001] direction ensured that bulk crystal structure is preserved.  Hence, the periodicity of all NW models corresponds to the $c$ lattice parameter of the bulk hexagonal unit cell, i.e.~28.65~$\mathrm{\AA}$, equivalent to that in Ref.~\onlinecite{Kobayashi2011b}. In total six different NW models were considered, where the NW perimeter $P_{NW}$ varied in the range 8.2~nm $\leq P_{NW} \leq$ 30.4~nm. 

For nanoribbons (Fig.~\ref{fig:Bi2Se3_HexNW_NR_Models}c,d), the axial direction is the [2$\bar{1}\bar{1}$0] direction, which corresponds to  the $a$ lattice vector. This is orthogonal to the $c$ direction along which QLs stack, and is also the commonly observed experimental growth direction of Bi$_{2}$Se$_{3}$ NRs.\cite{Zou2014,Kong2009} The constructed NR models have two distinct structural degrees of freedom. The first being the thickness, which is dictated by the number of QLs stacked parallel to $c$, and the second being the width, which defines the dimension orthogonal to the thickness and the periodicity direction (Fig.~\ref{fig:Bi2Se3_HexNW_NR_Models}c). An example of a 3QL thick and $\approx$6~nm wide NR is shown in Fig.~\ref{fig:Bi2Se3_HexNW_NR_Models}c,d.
The direction of periodicity of NRs is parallel to the $a$ direction of the Bi$_{2}$Se$_{3}$ bulk hexagonal unit cell, i.e.~$a =4.11$~$\mathrm{\AA}$, equal to that in Ref.~\onlinecite{Kobayashi2011b}.
The NR models exhibit two different sets of facets. One represents the equivalent top and bottom facets, normal to the $c$ axis (Fig.~\ref{fig:Bi2Se3_HexNW_NR_Models}c), which correspond to the widely studied (0001) surface. The other set is the equivalent side facets, aligned in the orthogonal direction, parallel to the (10$\bar{1}$0) surface
and whose termination corresponds to the [2$\bar{1}\bar{1}$0] QL edge termination, as discussed above for hexagonal NWs, with a side surface angle $\theta  = 57^{\circ}$.\cite{Virk2016} The atomic structure of these facets is schematically shown in Fig.~\ref{fig:Bi2Se3_HexNW_NR_Models}c. 
Importantly, this is the only high-index surface that, to our knowledge, has been investigated both theoretically,\cite{Moon2011,Virk2016} and experimentally.\cite{Xu2013} 
However, we do not exclude the possible presence of other surface orientations and terminations in TI nanostructures, such as the ones investigated in Ref.~\onlinecite{Villanova16}.
In our study, we varied the NR width between 6 and 19~nm. For each width, band structure calculations were performed on NR supercells where the thickness was also varied between 5QL and 9QL (5--9~nm). In total, twenty five different NR models were considered, where the range of NR perimeters $P_{NR}$ was in the range 23.8~nm $\leq P_{NR} \leq$ 59.2~nm. 

\section{Results and discussion}

\subsection{A simple model}

We begin our discussion by introducing a simple continuum model that was successfully used to describe the electronic structure of 1D TI nanostructures.\cite{Zhang09,Egger10,Zhang2010} With the help of this model, we discuss how the spin helicity of the 2D Dirac cone surface-state band is manifested in the quantized sub-bands of the considered 1D nanostructures. This general picture will then be corroborated by the results of TB calculations of Bi$_{2}$Se$_{3}$ hexagonal nanowires and 
nanoribbons.

\begin{figure}
\centering
  \includegraphics[width=0.5\textwidth]{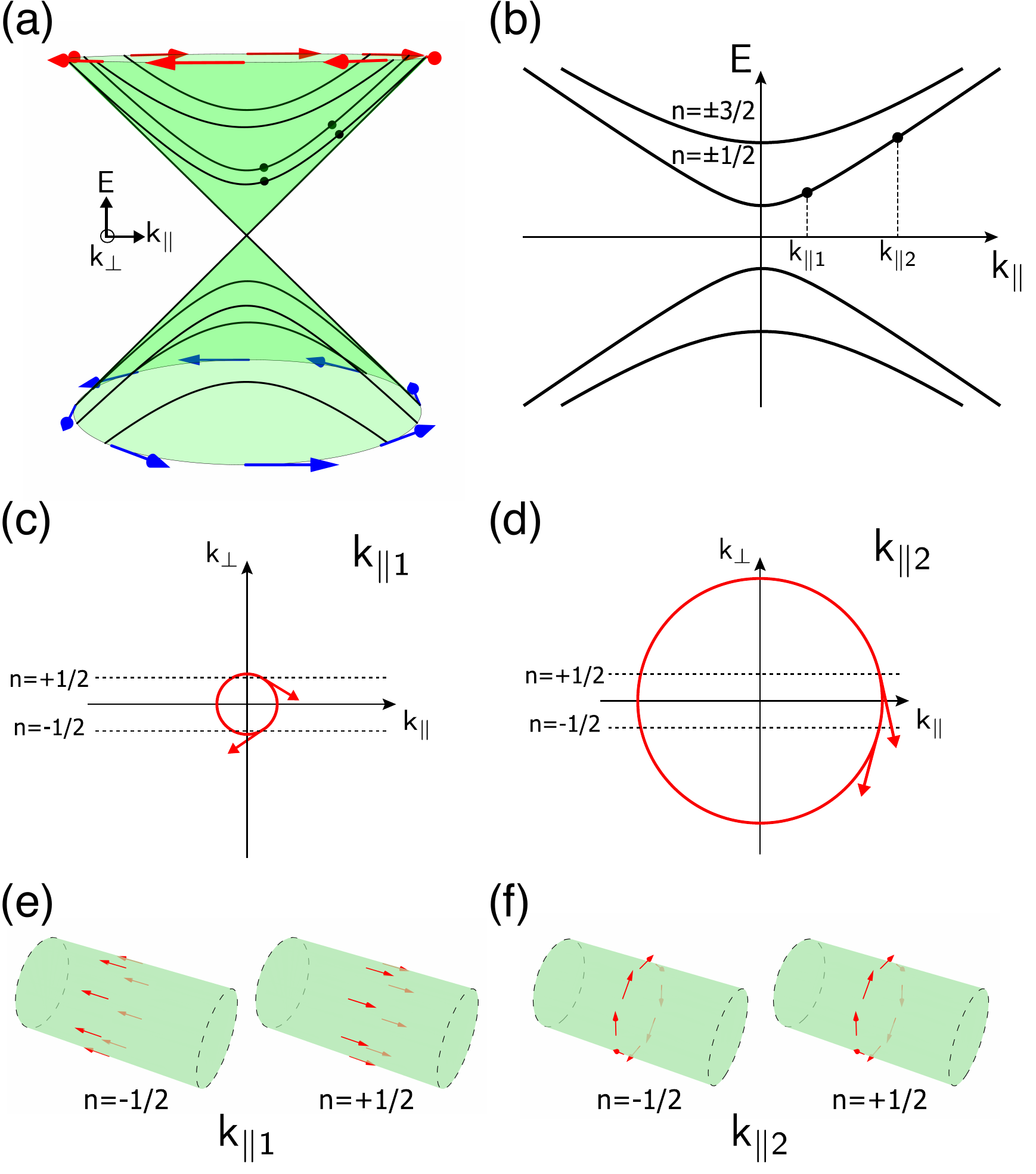}
  \caption{
(a) Schematic drawing of the 2D surface-state Dirac cone showing quantized 1D sub-bands as distinct cuts through the conical dispersion. Red and blue arrows around the conduction and valence band constant-energy contours show left- and right-handed helicities, respectively. 
(b) Band structure of a 1D nanostructure with the dispersion described by Eq.~(\ref{1D_nano_analytic_mod_full}) and $k_{\parallel}$ being the momentum along the nanostructure axis. 
(c,d) Schematic drawing showing the alignment of 
  the spin polarization vector ${\mathbf P}({\mathbf k})$ (Eq.~(\ref{Spin_pol_vect_ii})) along the associated constant-energy contour of the surface Dirac cone for the $n= \pm \frac{1}{2}$ states at $k_{\parallel1}$ and $k_{\parallel2}$. 
(e,f) Schematic drawing of the resultant real-space spin textures for $n= \pm \frac{1}{2}$ states at $k_{\parallel1}$ and $k_{\parallel2}$. }
  \label{fig:Gen_Mod_1D_Nanostruct}
\end{figure}

In this continuum model, we assume that the topological states on a flat 2D surface are described by an isotropic Dirac cone as shown in Fig.~\ref{fig:Gen_Mod_1D_Nanostruct}a. 
For purely illustrative purposes we will portray cylindrical nanowires and assume the independence of the surface-state Dirac cone on the surface orientation. 
One of the momentum directions,  $k_{\parallel}$ corresponds to the axial direction of the 1D nanostructure, i.e. along the $c$ and $a$ lattice vectors for NWs (Fig.~\ref{fig:Bi2Se3_HexNW_NR_Models}a) and NRs (Fig.~\ref{fig:Bi2Se3_HexNW_NR_Models}c), respectively. Subsequently, the direction of momentum $k_{\perp}$ orthogonal to $k_{\parallel}$ corresponds to the circumferential momentum around the nanostructure perimeter. The dimensions of the nanostructures under investigation are such that the magnitude of the perimeter is markedly smaller than that of the electron mean free path \cite{Analytis2010a,He2012}. Thus, surface electrons are subjected to  circular boundary conditions around the perimeter and hence $k_{\perp}$ is quantized.

Alongside the quantization of $k_{\perp}$, spin-momentum locking of the SS Dirac cone has a major effect on the band dispersion of a 1D nanostructure. The spin helicity results in an electron acquiring a $\pi$ Berry phase due to the 2$\pi$ rotation of the electron spin around the perimeter.\cite{Zhang09}  Consequently, the dispersion of the 1D sub-bands can be described by the following expression
\begin{align}
E(k_{\parallel}) &= \pm \hbar \nu_{F} \sqrt{k_{\parallel}^2+k_{\perp}^2}, \label{1D_nano_analytic_mod}
\end{align}
where $\nu_{F}$ is the Fermi velocity and the quantized values of $k_{\perp}$ are given by
\begin{align}
k_{\perp} &= \frac{2\pi(l+\frac{1}{2})}{P} \label{k_perp_quant}
\end{align}
with $P$ being the nanostructure perimeter.\cite{Egger10}
Thus, the dispersion of 1D sub-bands can be written as
\begin{align}	
E(l,k_{\parallel}) &= \pm \hbar \nu_{F} \sqrt{k_{\parallel}^2+\frac{[2\pi(l+\frac{1}{2})]^{2}}{P^{2}}} \label{1D_nano_analytic_mod_full},
\end{align}
where $l=(0,\pm 1, \pm 2, \ldots)$ is an angular momentum quantum number and the half-integer term ($\frac{1}{2}$) stems from the $\pi$ Berry phase. Without loss of generality in the rest of our paper we will use a half-integer index
\begin{align}
n &= (l+\frac{1}{2}) = (\pm \frac{1}{2}, \pm \frac{3}{2}, \pm  \frac{5}{2}, \ldots) . \label{half_int_ang_moment}
\end{align} 

The quantized values of $k_{\perp}$ (Eq.~(\ref{k_perp_quant})) can be seen as distinct cuts across the 2D SS Dirac cone (Fig.~\ref{fig:Gen_Mod_1D_Nanostruct}a). The electronic spectrum of a 1D nanostructure (Fig.~\ref{fig:Gen_Mod_1D_Nanostruct}{b}) is gapped due to the half-integer shift from the $\pi$ Berry phase in Eq.~(\ref{1D_nano_analytic_mod_full}). Consequently, the electronic spectrum is characterized by a series of doubly degenerate discrete 1D sub-bands, which can be labeled by half-integer $n = (\pm \frac{1}{2}, \pm \frac{3}{2},\ldots)$ (Eq.~(\ref{half_int_ang_moment})), and have a dispersion described by Eq.~(\ref{1D_nano_analytic_mod_full}). An example of such a band structure is given in Fig.~\ref{fig:Gen_Mod_1D_Nanostruct}{b}.
Increasing the nanostructure perimeter results in the decrease of both the sub-band separation and the band gap, thus recovering the band structure of projected Dirac cone of the 2D topological SSs in the limit of infinite perimeter.

Spin-momentum locking of the 2D SSs is shown by the spin-polarization vector ${\mathbf P}({\mathbf k})$ along the constant energy contours encircling the Dirac point, with
\begin{align}
{\mathbf P}\left ({\mathbf k}\right) &= \frac{2}{\hbar} \times \left [ \left \langle S_{x}({\mathbf k}) \right \rangle, \left \langle S_{y}({\mathbf k}) \right \rangle, \left \langle S_{z}({\mathbf k}) \right \rangle \right], \label{Spin_pol_vect_ii}
\end{align}
and the expectation value of the spin operators given by
\begin{align}
\left \langle S_{\alpha}({\mathbf k}) \right \rangle &= \frac{\hbar}{2} \left\langle \psi ({\mathbf k}) \lvert \sigma_{\alpha} \rvert \psi ({\mathbf k}) \right \rangle \quad \left( \alpha = x,y,z \right), \label{expec_val_spin_operators}
\end{align}
where $\psi ({\mathbf k})$ are the two-component spinor wavefunctions, and $\sigma_{\alpha}$ the corresponding Pauli matrices. For a 2D surface, defined by momenta $k_{\parallel}$ and $k_{\perp}$, spin-momentum locking is evinced by the spin-polarization vector pointing along the  $({\mathbf k} \wedge {\mathbf e}_z)$ direction.\cite{Hsieh2009,Liu2010,Zhang2010a,Yazyev2010} This gives rise to a helicity that is left (right) handed for the conduction (valence) band, as denoted by the red (blue) arrows in Fig.~\ref{fig:Gen_Mod_1D_Nanostruct}a. In Bi$_{2}$Se$_{3}$, the presence of strong spin-orbit interactions implies that the magnitude of ${\mathbf P}({\mathbf k})$ is reduced from a maximum value of 1 (i.e. 100\% spin polarization).\cite{Yazyev2010} For example, in the case of the Bi$_{2}$Se$_{3}$ (0001) surface, the magnitude of spin polarization is determined to be 0.5--0.6.\cite{Yazyev2010} A similarly reduced value was also found for the high-index surfaces defined by the stoichiometric [2$\bar{1}\bar{1}$0] QL edge termination.\cite{Virk2016} This is of particular relevance given that in our models the surfaces are composed of these two stoichiometric terminations, as discussed in Section~\ref{Meth}.

A direct correspondence between the spin helicity of the 2D surface Dirac cone and its manifestation in the quantized 1D nanostructure sub-bands can be established. For the  lowest-energy degenerate sub-band we subsequently choose a pair of degenerate states, labeled by their respective $n=\pm \frac{1}{2}$ values, at two different momenta $k_{\parallel 1}$ and $k_{\parallel 2}$  as shown in Fig.~\ref{fig:Gen_Mod_1D_Nanostruct}b. The $n=\pm \frac{1}{2}$ states at $k_{\parallel 1}$ are positioned close to $k_{\parallel} = 0$, while $k_{\parallel 2} \approx k$. 

The states at $k_{\parallel 1}$ are situated towards the centre of the $n=\pm \frac{1}{2}$ cuts across the 2D surface Dirac cone (Fig.~\ref{fig:Gen_Mod_1D_Nanostruct}a). The spin-polarization vector ${\mathbf P}({\mathbf k})$ (Eq.~(\ref{Spin_pol_vect_ii})) along the associated constant-energy contour is therefore aligned almost completely parallel to $+k_{\parallel}$ ($-k_{\parallel}$) for the $n=+\frac{1}{2}$ ($n=-\frac{1}{2}$) state (Fig.~\ref{fig:Gen_Mod_1D_Nanostruct}c), with a smaller component aligned along $k_{\perp}$. The reversal of the alignment of ${\mathbf P}({\mathbf k})$ along $k_{\parallel}$ being a direct consequence of the spin helicity of the surface Dirac cone. In real space, this should give rise to a nanostructure spin texture that is oriented along opposing directions of the nanostructure axis for each $n=+\frac{1}{2}$ and $n=-\frac{1}{2}$ degenerate state at $k_{\parallel 1}$, as shown schematically in Fig.~\ref{fig:Gen_Mod_1D_Nanostruct}e. Conversely, ${\mathbf P}({\mathbf k})$ is now aligned predominantly parallel to $-k_{\perp}$ for both $n=\pm \frac{1}{2}$ states at $k_{\parallel 2}$ (Fig.~\ref{fig:Gen_Mod_1D_Nanostruct}{d}), with a smaller component aligned along $k_{\parallel}$. In real space, at $k_{\parallel 2}$, this should give rise to a nanostructure spin texture that is oriented in the plane perpendicular to its axis along the direction tangential to its circumference, and rotates in the same direction around the perimeter for both $n=\pm \frac{1}{2}$  states, as shown schematically in Fig.~\ref{fig:Gen_Mod_1D_Nanostruct}f. We note that in the first case ($k_{\parallel 1}$) the spin polarization of the $n=\pm \frac{1}{2}$ degenerate states is close to antiparallel, while it's nearly parallel in the second case ($k_{\parallel 2}$).

Below, we will be referring to this simple model when discussing the results of our numerical calculations of the atomistic models of 1D NWs and NRs of Bi$_{2}$Se$_{3}$.

\subsection{Hexagonal nanowires}

\begin{figure}[b]
  \centering
  \includegraphics[width=0.45\textwidth]{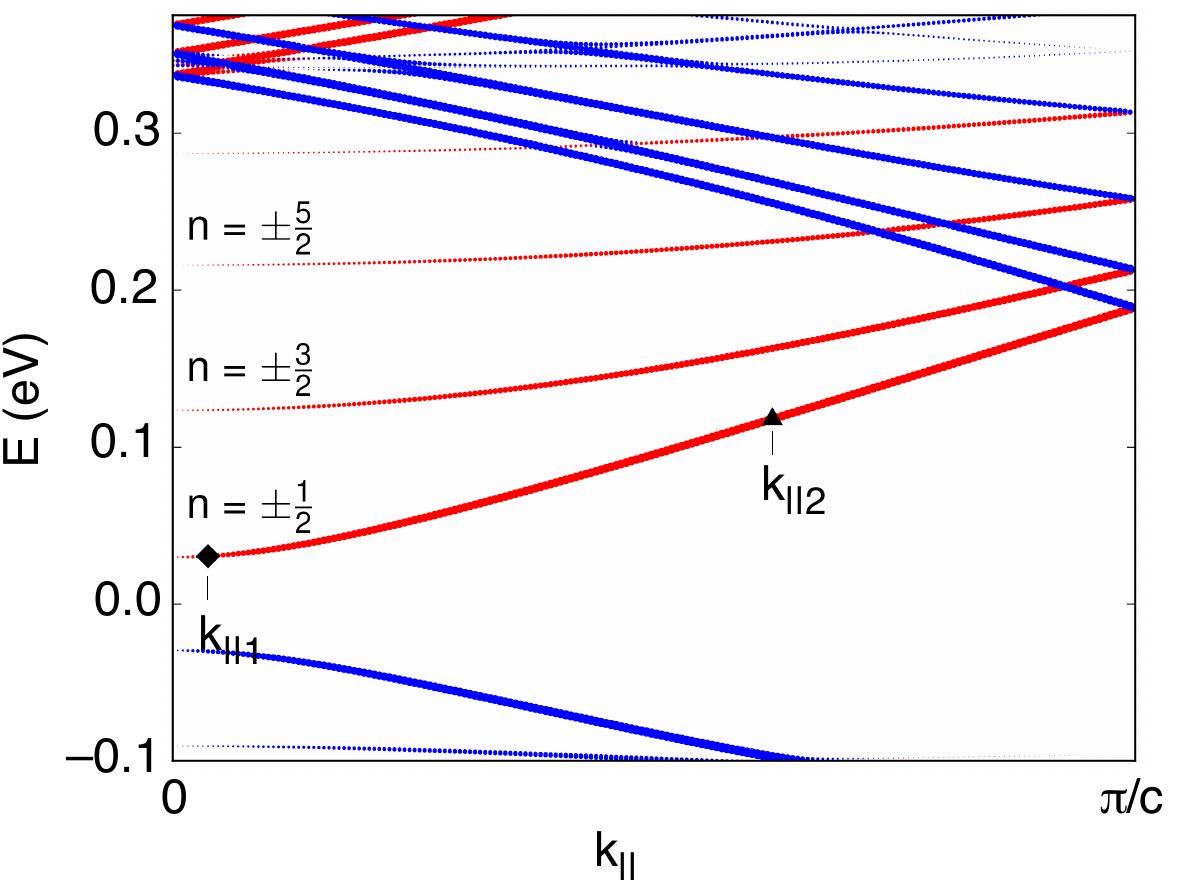}
  \caption{Tight-binding band structure of Bi$_2$Se$_3$ hexagonal nanowire with $P_{NW} = 30.4$~nm. 
The bands are colored according the sign of $h^{n}(k_{\parallel})$, while the symbol size reflects its magnitude.
The two states discussed in the text are labeled. Zero energy corresponds to the middle of the band gap.}
  \label{fig:Hex_NW_Band}
\end{figure}

Band structures for the investigated range of hexagonal NW models (8.2~nm $\leq P_{NW} \leq$ 30.4~nm) were computed using a tight-binding model as described in Section~\ref{Meth}. A representative band structure for the largest NW model ($P_{NW} = 30.4$~nm) is shown in Fig.~\ref{fig:Hex_NW_Band}. The band structure is characterized by the presence of discrete 1D sub-bands with a hyperbolic dispersion, as expected from the quantization along $k_{\perp}$. We find that a finite direct band gap ($E_{NW}$), located at $k_{\parallel}=0$, and doubly degenerate bands characterize the electronic spectrum of all NWs. Unlike in the simple model discussed above (Fig.~\ref{fig:Gen_Mod_1D_Nanostruct}b), the realistic band structure in Fig.~\ref{fig:Hex_NW_Band} shows the expected electron-hole asymmetry. Below, we will limit our discussion to sub-bands derived from the electron part of the Dirac cone ($E > 0$) and to the narrow energy range in which bulk states of Bi$_2$Se$_3$ are irrelevant. 

\begin{figure}
  \centering
  \includegraphics[width=0.4\textwidth]{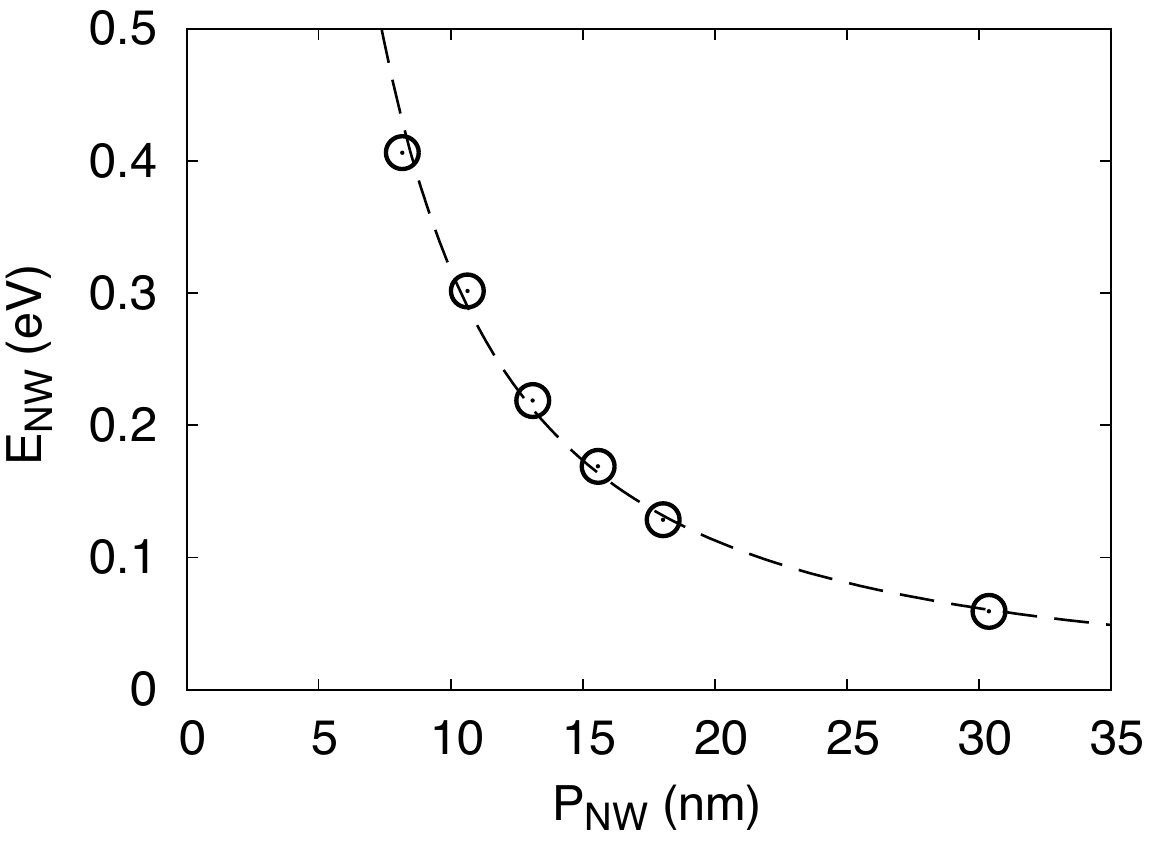}
  \caption{Calculated band gaps of Bi$_2$Se$_3$ nanowires $E_{NW}$  as a function of their perimeter $P_{NW}$. Numerical fit to $E_{NW} \propto P_{NW}^{-1.5}$ is given by dashed line.
}
  \label{fig:Hex_NWs_E_vs_perim}
\end{figure}

The dependence of band gap $E_{NW}$ on NW perimeter $P_{NW}$ is displayed in Fig.~\ref{fig:Hex_NWs_E_vs_perim}. As expected from the general model discussed above, the band gap and the inter-level spacing decrease with increasing $P_{NW}$. As the NW perimeter increases, $E_{NW}$ tends to 0, approaching the limit of the 2D surface in which a crossing of two linearly dispersing bands is recovered. From Eq.~(\ref{1D_nano_analytic_mod_full}) one would expect the $E_{NW} \propto P_{NW}^{-1}$ dependence. However, we find a relationship very close to $E_{NW} \propto P_{NW}^{-1.5}$  as clearly seen in Fig.~\ref{fig:Hex_NWs_E_vs_perim}. We believe that this discrepancy is due to the fact that the penetration depth of topological SSs is comparable to the NW width in the investigated range of NWs perimeters. For narrow NWs, the hybridization between the SSs across the NW bulk may additionally contribute to band gap opening. 

We now concentrate on the results obtained for the largest investigated NW ($P_{NW} = 30.4$~nm), whose band structure is shown in Fig.~\ref{fig:Hex_NW_Band}.
Similarly to what was done for the general model above, we define a pair of $n=\pm \frac{1}{2}$ degenerate states, at two different momenta, $k_{\parallel 1} \approx 0$ and a larger $k_{\parallel 2}$. In order to analyze the real-space spin textures of these states we determine the real-space local spin density ${\mathbf S}_{i}$ computed from the expectation values of the spin operators projected onto atomic sites ${\mathbf r}_{i}$. At a particular momentum $k_{\parallel}$ and energy $E$ the momentum-resolved local spin density for a given state $n$ is
\begin{align}
{\mathbf S}^{n}_{i} \left( k_{\parallel},{\mathbf r}_{i} \right) &=  \frac{\hbar}{2} \left\langle \psi \left(k_{\parallel}\right) \lvert \sigma_{\alpha} \otimes P_{i} \rvert  \psi\left(k_{\parallel}\right) \right\rangle \quad \left( \alpha = x,y,z \right), \label{loc_spin_density}
\end{align}
where $P_{i}$ is a projector onto atom $i$ defined as $P_{i} = \sum_{\lambda\sigma} \lvert i \lambda\sigma\rangle \langle i \lambda\sigma \rvert$, with $\lambda$ being the orbital index and $\sigma$ the spin. 

\begin{figure*}
  \centering
  \includegraphics[width=0.6\textwidth]{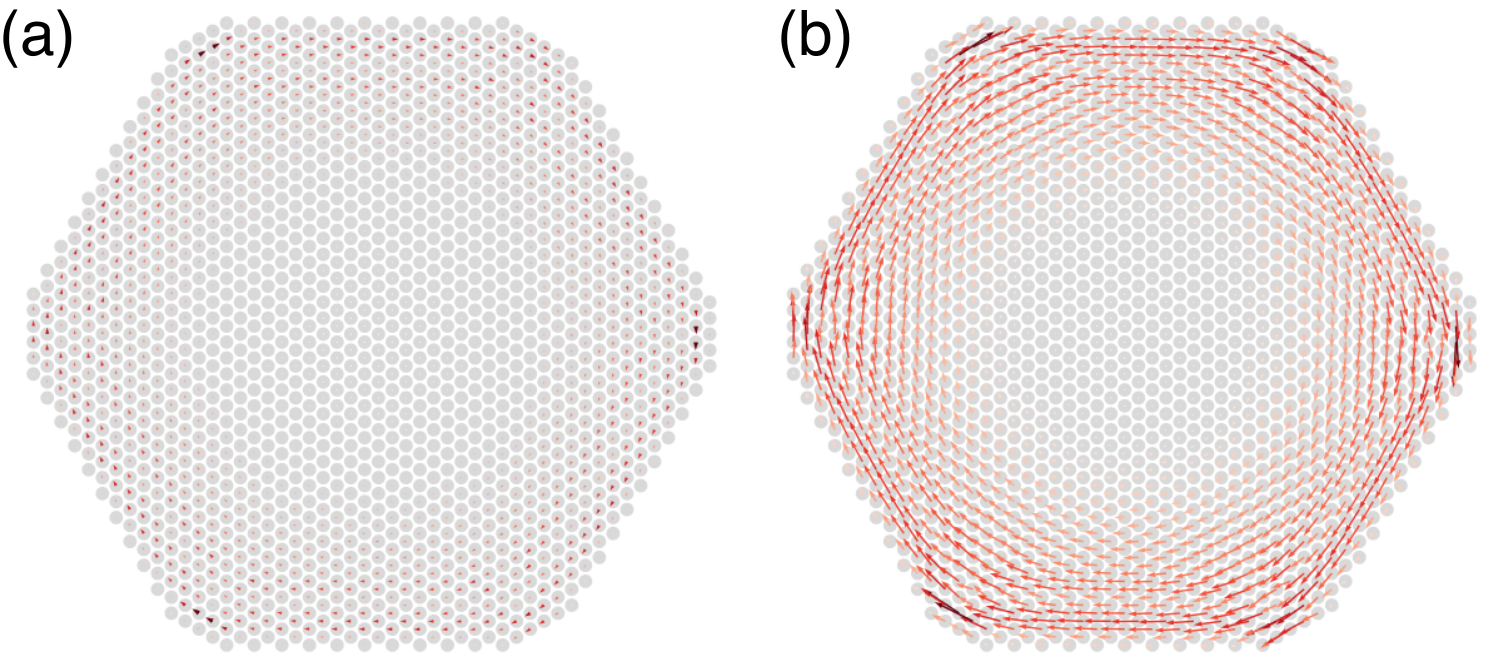}
  \caption{Spin textures of a Bi$_{2}$Se$_{3}$ hexagonal nanowire ($P_{NW}=30.4$ nm) in the cross-sectional view for the $n=\pm \frac{1}{2}$ states at (a) $k_{\parallel1}$ and (b) $k_{\parallel2}$. The arrows reflect the orientation and magnitude of the total local spin density ${\mathbf S}^{tot}_{i}$. Grey circles indicate the positions of Bi and Se atoms. }
\label{fig:NW_spin_texture}
\end{figure*}

A consequence of the reduced dimensionality of the NW and NR morphologies is that the wavefunction for each $n=\pm \frac{1}{2}$ degenerate state, at either at $k_{\parallel 1}$ or $k_{\parallel 2}$, is localized on the surface of a nanostructure around its entire perimeter. This contrasts to the case of 2D surfaces where calculations  are performed on models in slab geometries. For slabs of sufficient thickness the degenerate surface states of opposite spin helicities are localized at opposite surfaces.\cite{Zhang2010a,Yazyev2010}
Therefore, in order to discuss the real-space spin properties of 1D TI nanostructures we consider the sum ${\mathbf S}^{tot}_{i}$ of the local spin densities ${\mathbf S}^{n}_{i}$ for each of the two $n=\pm \frac{1}{2}$ degenerate states
\begin{align}
{\mathbf S}^{tot}_{i} &= {\mathbf S}^{+\frac{1}{2}}_{i} + {\mathbf S}^{-\frac{1}{2}}_{i}. \label{sum_loc_spin_density}
\end{align}
This total local spin density ${\mathbf S}^{tot}_{i}$ is plotted in Fig.~\ref{fig:NW_spin_texture} as a function of atomic position ${\mathbf r}_{i}$ for the largest considered NW ($P_{NW} \sim 30.4$ nm). The localization of ${\mathbf S}^{tot}_{i}$, at both $k_{\parallel 1}$ (Fig.~\ref{fig:NW_spin_texture}a) and $k_{\parallel 2}$ (Fig.~\ref{fig:NW_spin_texture}{b}), around the NW perimeter is evident, thereby confirming the surface-state origin of the discussed $n=\pm \frac{1}{2}$ states.

The difference in spin textures observed at $k_{\parallel 1}$ and $k_{\parallel 2}$ can be explained as follows.
At $k_{\parallel 1} \approx 0$, the spin-polarization vector ${\mathbf P}({\mathbf k})$ of the surface Dirac cone is oriented almost entirely along $+k_{\parallel}$ ($-k_{\parallel}$) for the $n=+\frac{1}{2}$ ($n=-\frac{1}{2}$) state (Fig.~\ref{fig:Gen_Mod_1D_Nanostruct}{c}).
Therefore, the major component of the NW local spin density ${\mathbf S}^{n}_{i}$ is oriented in opposing directions of the NW axis for each $n=\pm \frac{1}{2}$ state, resulting in spin textures analogous to the schematic drawing in Fig.~\ref{fig:Gen_Mod_1D_Nanostruct}{e}. In the total local spin density ${\mathbf S}^{tot}_{i}$ these major components consequently cancel each other. This leaves a minor component in the direction tangential to the NW circumference and an overall low degree of spin polarization, as shown in Fig.~\ref{fig:NW_spin_texture}{a}.
At a larger momentum $k_{\parallel 2}$, ${\mathbf P}({\mathbf k})$ is almost parallel to $-k_{\perp}$ for both $n=\pm \frac{1}{2}$ states (Fig.~\ref{fig:Gen_Mod_1D_Nanostruct}{d}). This is manifested in the major component of ${\mathbf S}^{n}_{i}$ being oriented in the direction tangential to the NW circumference, and rotating around the same direction of the perimeter for both $n=\pm \frac{1}{2}$ states. The resulting spin textures are analogous to those schematically shown in Fig.~\ref{fig:Gen_Mod_1D_Nanostruct}f. Upon computing the total local spin density ${\mathbf S}^{tot}_{i}$  these dominant components of ${\mathbf S}^{+\frac{1}{2}}_{i}$ and ${\mathbf  S}^{-\frac{1}{2}}_{i}$ sum constructively, whilst the minor components aligned along opposing directions of the NW axis cancel each other. The resultant spin texture (Fig.~\ref{fig:NW_spin_texture}{b}) lies entirely in the plane perpendicular to the NW axis, has a pronounced clockwise 2$\pi$ rotation of ${\mathbf S}^{tot}_{i}$ around the NW perimeter, and an overall larger degree of spin polarization. The handedness of this rotation is directly related to the left-handed helicity observed for the 2D TI conduction band SS (red arrows in Fig.~\ref{fig:Gen_Mod_1D_Nanostruct}{a}). This establishes the direct correspondence between the spin helicity of the 2D surface Dirac cone and the real space spin texture of the quantized sub-bands of the 1D TI NW.

To quantify the 2$\pi$ rotation of the local spin density around the NW perimeter, for a given sub-band index $n$ at a particular momentum $k_{\parallel}$, we calculate the following quantity
\begin{align}
h^{n}(k_{\parallel}) &= \sum_{i} \left({\mathbf S}^{n}_{i} \left(k_{\parallel},{\mathbf r}_{i}\right) \times\frac{{\mathbf r}_{i}}{\lvert r_{i} \rvert} \right), \label{Hel_mag} 
\end{align}
where ${\mathbf S}^{n}_{i} \left(k_{\parallel},{\mathbf r}_{i}\right)$ is the momentum-resolved local spin density on a given atom $i$, as defined by Eq.~(\ref{loc_spin_density}), whilst ${\mathbf r}_{i}$ is the atomic position vector from the 1D nanostructure axis. An ideal 2$\pi$ rotation of the local spin density ${\mathbf S}^{n}_{i}$ around the NW perimeter, with the spin aligned tangential to the circumference of the NW, would be reflected in a maximum value of $h^{n}$. Under the assumption of a 100$\%$ spin polarization, this value would equal $h^{n}=\pm1$ for a clockwise and anti-clockwise rotation, respectively. However, for Bi$_{2}$Se$_{3}$ NWs, the maximum possible absolute value of $h^{n}$ is lower than 1, as spin polarization is less than 100$\%$ due to the strong spin-orbit interaction.\cite{Yazyev2010}  Subsequently, the largely in plane 2$\pi$ rotation of ${\mathbf S}^{n}_{i}$ around the NW perimeter at $k_{\parallel 2}$ corresponds to a value of $h^{n}(k_{\parallel 2}) = 0.64$ for the $n=\pm \frac{1}{2}$ states. Conversely, the fact that ${\mathbf S}^{n}_{i}$ tends to align along the NW axis at $k_{\parallel 1}$ is reflected in a significantly lower value $h^{n}(k_{\parallel 1}) = 0.14$ for the same $n=\pm \frac{1}{2}$ states.

Further correspondence between the spin helicity of the 2D surface states and the quantized 1D NW sub-bands is given by analyzing the dependence of $h^{n}(k_{\parallel})$ on sub-band index $n$. It is  evident from Fig.~\ref{fig:Hex_NW_Band} that for a given value of $k_{\parallel}$ the magnitude of $h^{n}$ decreases upon increasing $n$ (as well as energy $E$ in the case of conduction band). This is because for the respective higher energy sub-bands (e.g. $n = \pm \frac{3}{2}, \pm \frac{5}{2}$), the alignment of the spin-polarization vector ${\mathbf P}({\mathbf k})$ of the surface Dirac cone along $k_{\perp}$ ($k_{\parallel}$) decreases (increases) with increasing $n$. Consequently, in real space this reduces the tangential component of ${\mathbf S}^{n}_{i}$ and increases its alignment along the NW axis, thereby reducing the magnitude of $h^{n}(k_{\parallel})$.

\begin{figure}[h]
  \centering
  \includegraphics[width=0.45\textwidth]{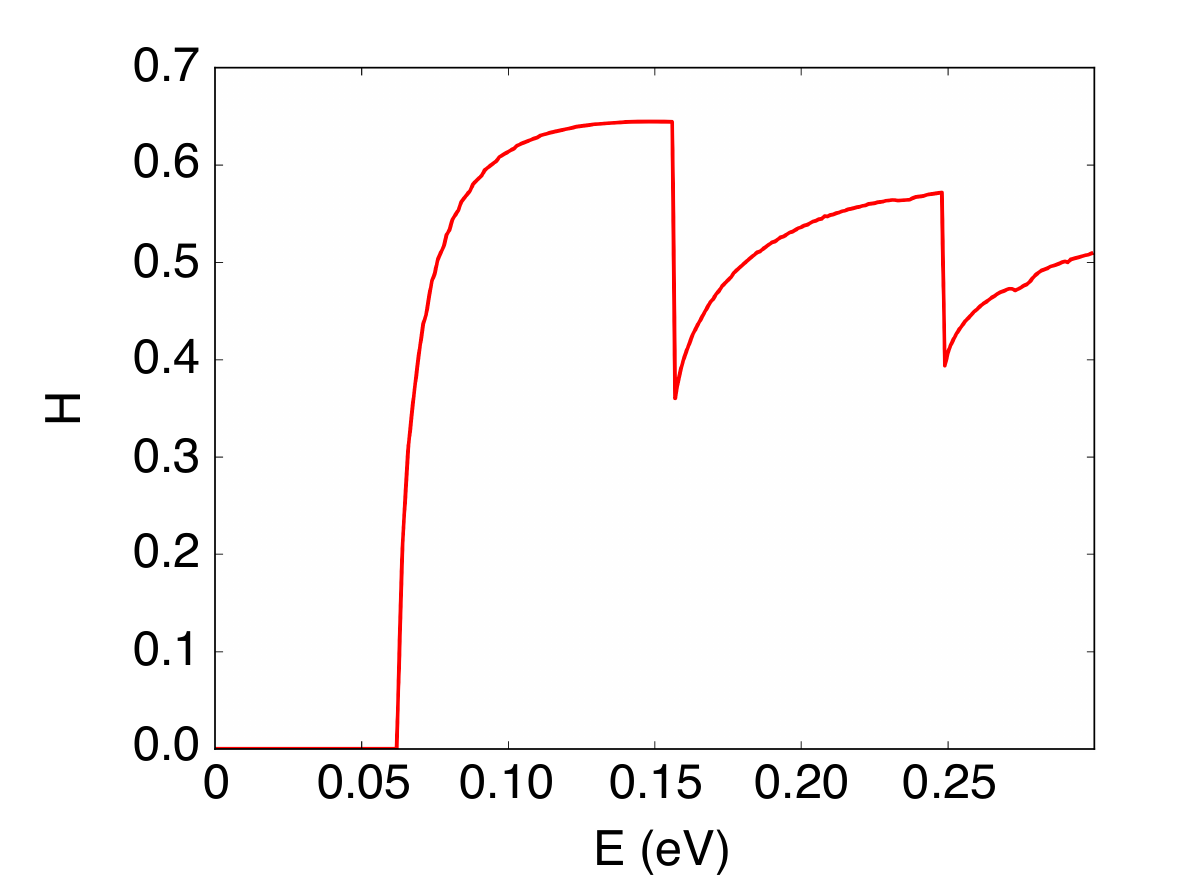}
  \caption{Tangential spin polarization density of forward propagating states $H(E)$ as a function of energy $E$ calculated for the $P_{NW}=30.4$~nm Bi$_2$Se$_3$ hexagonal nanowire. Zero energy corresponds to valence band maximum.}
\label{fig:spin_pol_dens} 
\end{figure}

We anticipate that this peculiar behavior of $h^{n}(k_{\parallel})$ may give rise to spin transport properties of potential interest for practical applications. In order to explore these properties, we define the tangential spin polarization density of the forward propagation states as 
\begin{align}
H(E) &= \frac{\int_0^{\pi/c} \sum_{n} h^{n}(k_{\parallel}) \delta (E-E_{n,k_{\parallel}}) dk_{\parallel}}{\int_0^{\pi/c}\sum_{n} \delta (E-E_{n,k_{\parallel}})dk_{\parallel}}, \label{loc_spin_pol}
\end{align}
where the denominator defines the number of available transport channels at a given energy $E$, and the numerator is the sum of $h^{n}$ with respect to those available transport channels. 
The clear-cut oscillatory dependence of $H(E)$ shown in Fig.~\ref{fig:spin_pol_dens} for the $P_{NW} = 30.4$~nm NW reflects the quantization along $k_{\perp}$, and the variation of $h^{n}(k_{\parallel})$. 
Each step-like feature crossed upon increasing the charge-carrier energy corresponds to adding a pair of conductance channels due to quantized sub-bands. However, the small values of $h^{n}(k_{\parallel})$ at the sub-band edges ({\it i.e.} at $k \approx 0$) result in sharp drops of $H(E)$  upon including these new conductance channels. 
The non-zero values of $H(E)$ imply that a charge current induces local spin density on the NW surface, even though the net spin polarization is zero due to their winding spin texture of the discussed states in TI NWs.
Moreover, in contrast to the case of flat 2D surfaces of TIs where the dependence of spin polarization on charge-carrier energy is rather weak,\cite{Yazyev2010} 1D NWs display the discussed oscillatory behavior stemming from the reduced dimensionality. We propose that such a peculiar strong dependence can be exploited in spintronic devices. TI NWs with locally placed contacts, {\it e.g.} covering only one side of the surface, can be used for injecting spin-polarized charge carriers into other materials, while the degree of spin-polarization can be controlled through gating.

\subsection{Nanoribbons}

\begin{figure}[b]
  \centering
  \includegraphics[width=0.45\textwidth]{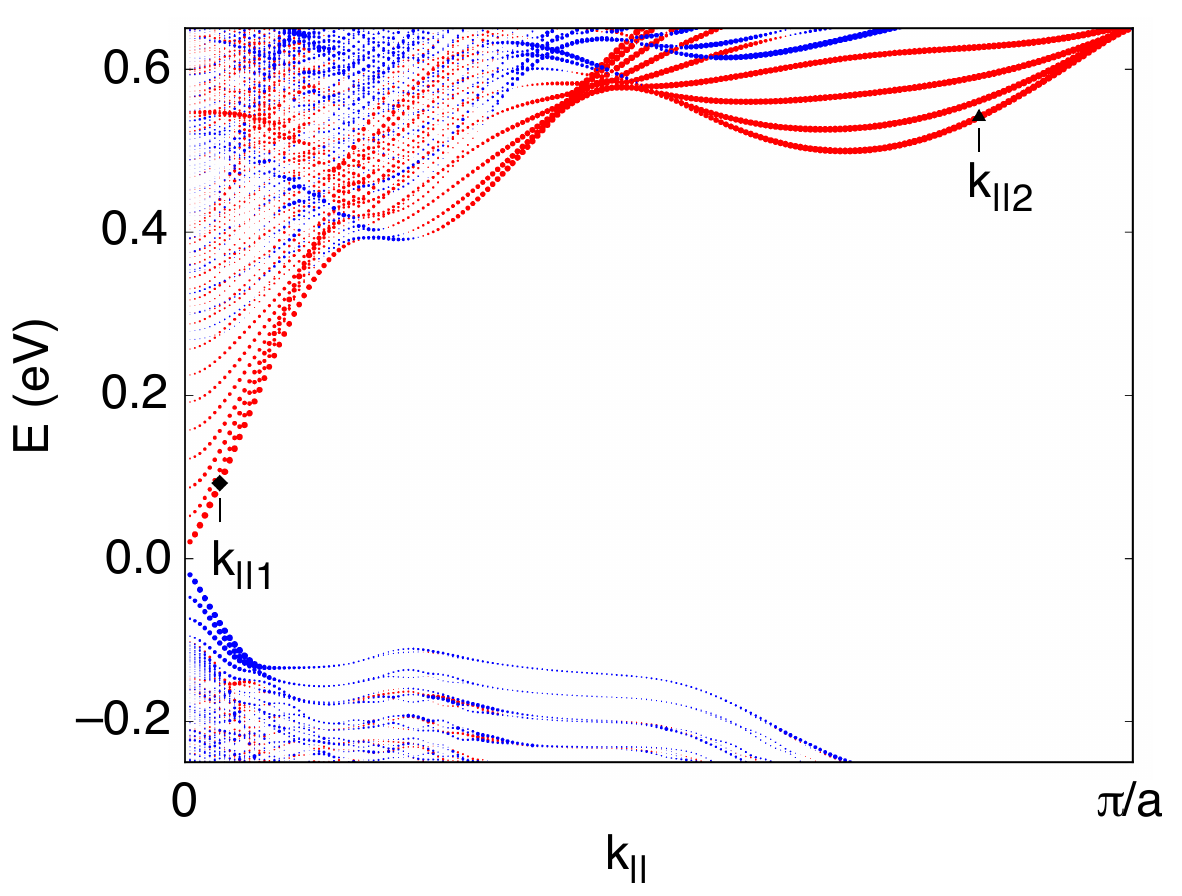}
  \caption{
Tight-binding band structure of Bi$_2$Se$_3$ nanoribbon of 19~nm width and 9QL thickness ($P_{NR} = 59.2$~nm). 
The bands are colored according the sign of $h^{n}(k_{\parallel})$, while the symbol size reflects its magnitude.
The two states discussed in the text are labeled. Zero energy corresponds to the middle of the band gap.}
  \label{fig:NR_Band}
\end{figure}

Using the same TB approach as for hexagonal NWs, we investigated a number of nanoribbon models of different width and thickness, which corresponds to the range of perimeters 23.8~nm $\leq P_{NR} \leq$ 59.2~nm (see Section~\ref{Meth}). Overall, many qualitative trends observed for hexagonal NWs are found for NRs too. For example, the band structures of all investigated NRs exhibit a finite direct band gap $E_{NR}$ defined by band extrema at $k_{\parallel}=0$, and that $E_{NR}$ decreases with increasing $P_{NR}$. Furthermore, in the energy range that corresponds to the bulk band gap the band structures of NRs are characterized by a series of degenerate, evenly spaced hyperbolic sub-bands and their spin helicities $h^n(k_{\parallel})$ show the same momentum dependence as in the case of NWs. Thus, we will focus our discussion on differences between these two nanostructure morphologies. The differences are in part due to more complex structures of NRs. Firstly, there are two distinct parameters defining the NR structure: the thickness along the stacking direction defined by lattice constant $c$ and the width. 
Secondly, the NR morphology implies the presence of two different facets for which the topological SS band dispersion is different. One type of facets (referred to as top and bottom facets) corresponds to the extensively investigated (0001) surface, which is characterized by an isotropic Dirac cone centered at the $\Gamma$ point of the surface Brillouin zone.\cite{Zhang2009,Xia2009,Hsieh2009,Chen2009} 
The other type, referred to as side facets, are equivalent to that of the high-index surface defined by the [2$\bar{1}\bar{1}$0] QL edge termination and a surface angle $\theta = 57^{\circ}$. This surface is characterized by an anisotropic Dirac cone centered at the $\Gamma$ point of the surface Brillouin zone and lower Fermi velocities.\cite{Moon2011,Virk2016}

Figure~\ref{fig:NR_Band} shows the TB band structure of the largest investigated NR model characterized by 19~nm width and 9QL thickness that corresponds to $P_{NR} = 59.2$~nm. At first glance, this band structure looks different from the one calculated for a hexagonal NW model (Fig.~\ref{fig:Hex_NW_Band}). However, these differences can largely be explained by the following two conditions. Firstly, the two nanostructure morphologies are characterized by different periodicities. The lattice constant of the NW models is defined the bulk lattice constant $c = 28.65$~\AA\, while that of the NR models is defined by $a = 4.11$~\AA. Therefore, the Brillouin zone boundary in the case of NR models is characterized by a significantly larger momentum compared to the NW models, $k_{\parallel} = 0.76$~$\mathrm{\AA}^{-1}$ as opposed to $k_{\parallel} = 0.11$~$\mathrm{\AA}^{-1}$, respectively. Secondly, in Fig.~\ref{fig:NR_Band} we show a larger energy range that covers sub-bands derived from the bulk electronic states of Bi$_2$Se$_3$. It's worth noting that some of these bulk-like states also show a 
pronounced spin helicity $h^n(k_{\parallel})$. Otherwise, sub-bands resulting from circumferential quantization of SS Dirac cone and their spin helicities $h^n(k_{\parallel})$ are similar for the discussed NW and NR models shown in Figs.~\ref{fig:Hex_NW_Band} and \ref{fig:NR_Band}, respectively.

\begin{figure}
  \centering
  \includegraphics[width=0.45\textwidth]{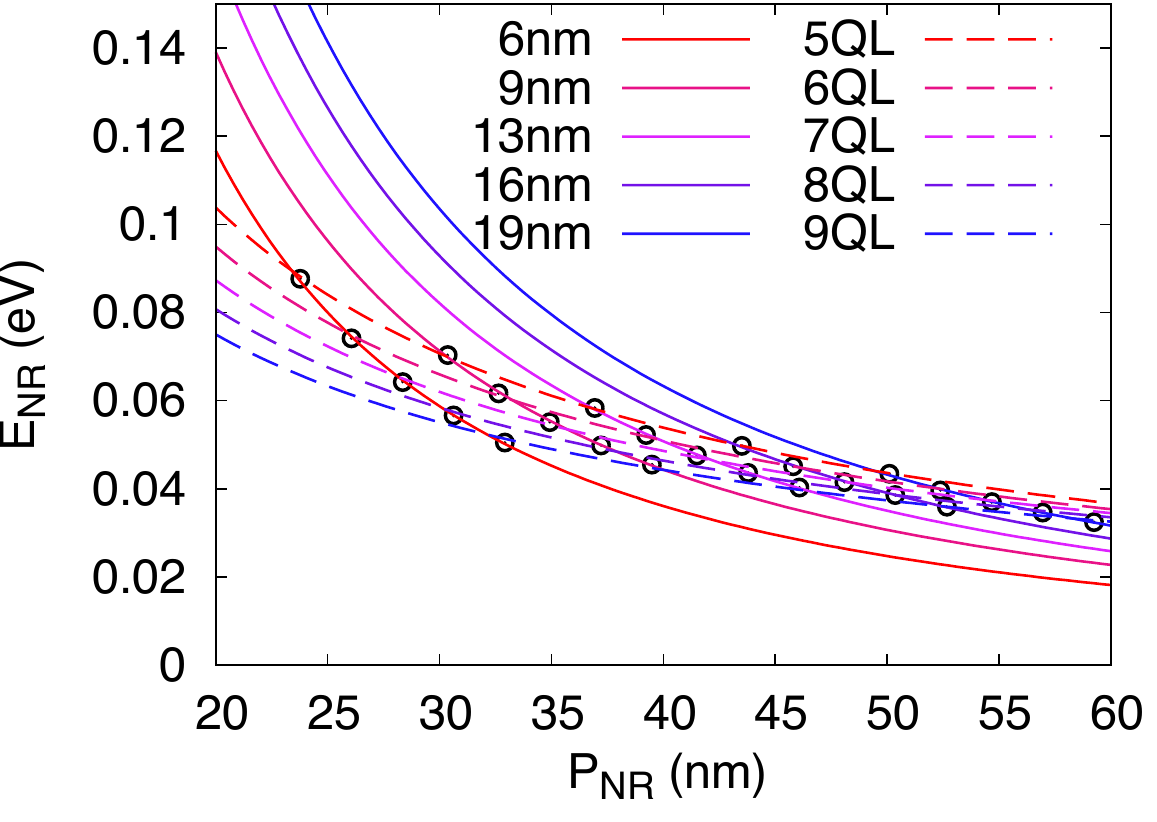}
  \caption{Calculated band gaps of Bi$_2$Se$_3$ nanoribbons $E_{NR}$  as a function of their perimeter $P_{NR}$.
Numerical fits for the models of same width (solid lines) and same thickness (dashed lines) are shown.}
  \label{fig:NRs_E_vs_perim}
\end{figure}

\begin{figure}
  \centering
  \includegraphics[width=0.5\textwidth]{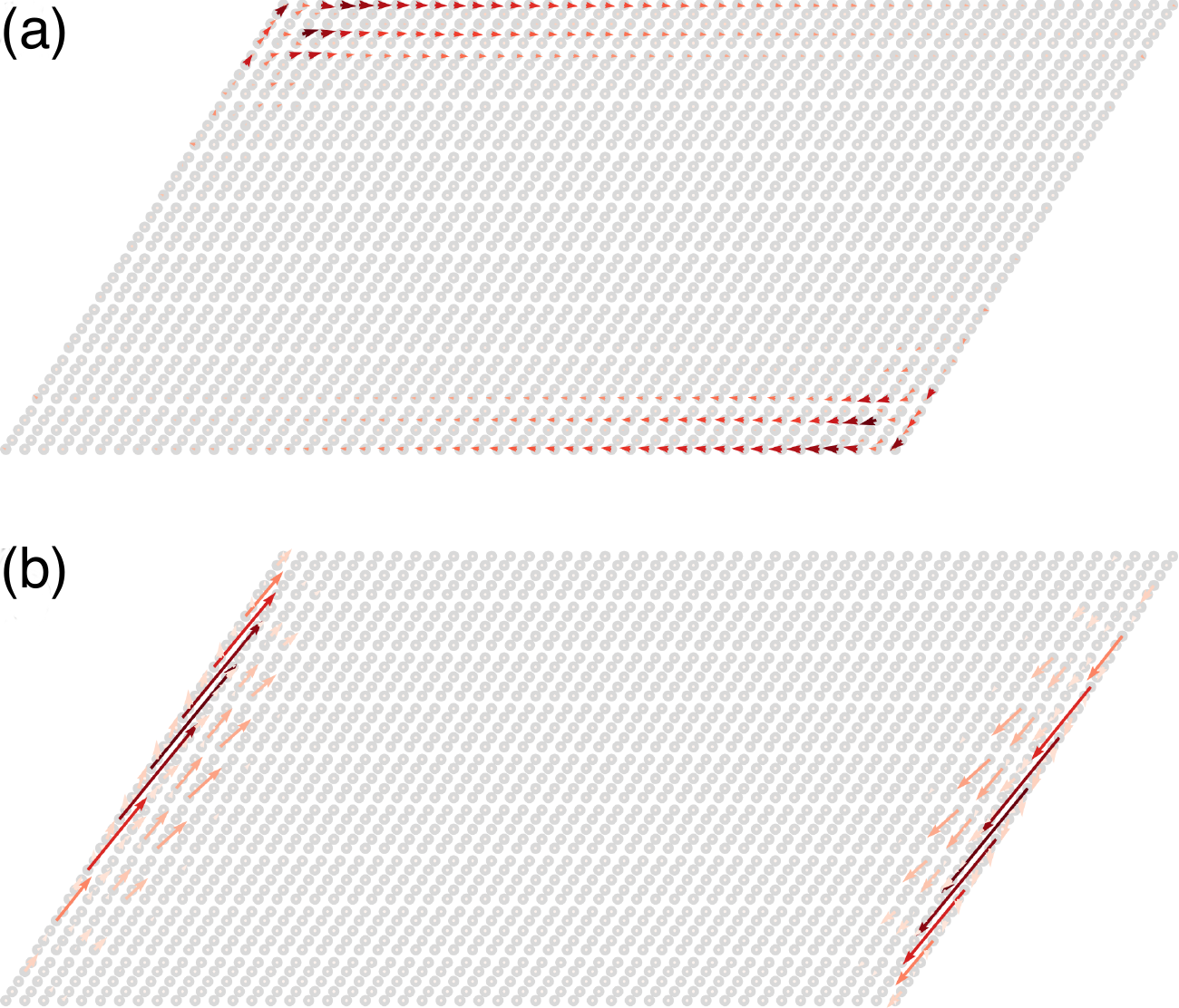}
  \caption{Spin textures of a Bi$_{2}$Se$_{3}$ nanorribon of 19~nm width and 9QL thickness ($P_{NR}=59.2$~nm) in the cross-sectional view for the $n=\pm \frac{1}{2}$ states at (a) $k_{\parallel1}$ and (b) $k_{\parallel2}$. The arrows reflect the orientation and magnitude of the total local spin density ${\mathbf S}^{tot}_{i}$. Grey circles indicate the positions of Bi and Se atoms.} \label{fig:NR_spin_texture}
\end{figure}

The presence of two structural degrees of freedom, represented by the two facet types, results in a more complex dependence of band gap $E_{NR}$ on NR perimeter $P_{NR}$ (Fig.~\ref{fig:NRs_E_vs_perim}). The values of $E_{NR}$ cannot be fitted by a single power-law dependence, but accurate fits can be obtained when NR width and thickness are considered independently.  
Firstly, for constant NR widths  (solid lines in Fig.~\ref{fig:NRs_E_vs_perim}) our calculations predict a scaling of $E_{NR} \sim P_{NR}^{-1.7}$ for all investigated values of NR width. This relationship is similar to that found for hexagonal NWs (Fig.~\ref{fig:Hex_NWs_E_vs_perim}). Conversely, for constant values of thickness (dashed lines in Fig.~\ref{fig:NRs_E_vs_perim}) the dependence of $E_{NR}$ on $P_{NR}$ becomes less strong as the thickness increases. This is reflected in an incremental reduction in the calculated power-law decay describing $E_{NR}(P_{NR})$, from $E_{NR} \propto P_{NR}^{-0.95}$ for 5QL-thick NRs (red dashed line in Fig.~\ref{fig:NRs_E_vs_perim}) to $E_{NR} \propto P_{NR}^{-0.75}$ for 9QL-thick NRs (blue dashed line in Fig.~\ref{fig:NRs_E_vs_perim}). 
The slower power-law decay upon the change of width can be explained by the fact that the surface states at the band edges the localized predominantly at the (0001) facets (Fig.~\ref{fig:NR_spin_texture}), and hence the band gap is to large extent defined by the SS hybridization across the NR thickness. 

Similarly to NW models, in Fig.~\ref{fig:NR_spin_texture} we plot real-space spin textures for two pairs of $n = \pm \frac{1}{2}$ degenerate states belonging to the lowest energy sub-band at two different momenta $k_{\parallel1}$ and $k_{\parallel2}$ in the discussed NR model.
The first momentum $k_{\parallel1}$ is close to $k_{\parallel} = 0$ and correspond to the state derived from the SS Dirac cone, while the investigated state at $k_{\parallel2}$ corresponds to a larger energy at which bulk-like states exist as well (see symbols in Fig.~\ref{fig:NR_Band}). The $n = \pm \frac{1}{2}$ states at both $k_{\parallel1}$ and  $k_{\parallel2}$ are non-uniformly distributed over the NR surface as a direct consequence of two structurally distinct facets. At $k_{\parallel1}$, the $n = \pm \frac{1}{2}$ states are localized at the facet formed by the (0001) surface (Fig.~\ref{fig:NR_spin_texture}a). The 2$\pi$ rotation is not evident because of such inhomogeneous distribution, but the spin-momentum locking is clearly manifested in the spin texture. The corresponding  $h^{n}(k_{\parallel})$ value is reduced as discussed for the case of hexagonal NWs. In contrast, at $k_{\parallel2}$, the $n = \pm \frac{1}{2}$ states are localized on the NR side facets defined by the stoichiometric  [2$\bar{1}\bar{1}$0] QL edge termination  (Fig.~\ref{fig:NR_spin_texture}a). The spin-momentum locking is also pronounced in this case and the corresponding value of $h^{n}(k_{\parallel})$ is larger. As observed for hexagonal NWs, at both momenta the ${\mathbf S}^{tot}_{i}$ vectors are in the tangential plane of the NR. We note also that the SSs in NRs appear to have a smaller penetration depth compared to the investigated models of hexagonal NWs (Fig.~\ref{fig:NW_spin_texture}). 
\\

\section{Conclusions}

In summary, we performed a numerical investigation of realistic models of experimentally observed one-dimensional nanowires and nanoribbons of the Bi$_2$Se$_3$ bulk topological insulator.
The low-energy sectors of the band structures are composed of evenly spaced degenerate sub-bands resulting from circumferential confinement of the 2D topological surface states that are described by the Dirac cone dispersion. The observed direct band gaps at $k_\parallel = 0$ are due to non-trivial $\pi$ Berry phase, and their magnitude decays as the nanostructure circumference increases. For nanowires, we find that the decay rate is faster than what one would expect from the simple circumferential confinement picture. A more complicated dependence was found for nanoribbons as two structural degrees of freedom and two distinct types of facets are present. The most intriguing finding is the 2$\pi$  rotation in the real-space spin texture of the discussed low-energy electronic states, and a clear dependence of the corresponding degree of spin polarization on momentum along the nanostructure. Both effects are rationalized within the same simple picture relying on the circumferential confinement of the topological surface states. The same  circumferential  quantization is also responsible for the oscillating behavior of the surface spin density induced by electric current along the nanostructure on the charge-carrier energy, which has clear implications for spintronic devices.

 
\begin{acknowledgments}
This work was supported by the Swiss National Science Foundation (grant No. PP00P2\_133552) and the ERC project ``TopoMat'' (grant No. 306504). The calculations were performed at the Swiss National Supercomputing Centre (CSCS) under projects s515 and s675.
\end{acknowledgments}


%

\end{document}